# The Application of AHP Model to Guide Decision Makers: A Case Study of E-Banking Security

Irfan Syamsuddin
State Polytechnic of Ujung Pandang
Makassar, Indonesia
e-mail: irfans@poliupg.ac.id

Junseok Hwang
Seoul National University
Seoul, Republic of Korea
e-mail: junhwang@snu.ac.kr

*Abstract*—Changes in technology have resulted in new ways for bankers to deliver their services to costumers. Electronic banking systems in various forms are the evidence of such advancement. However, information security threats also evolving along this trend. This paper proposes the application of Analytic Hierarchy Process (AHP) methodology to guide decision makers in banking industries to deal with information security policy. The model is structured according aspects of information security policy in conjunction with information security elements. We found that cultural aspect is valued on the top priority among other security aspects, while confidentiality is considered as the most important factor in terms of information security elements.

*Information security; policy; decision making; AHP*

## I. INTRODUCTION

Changes in technology have resulted in new ways for bankers to deliver their services to costumers. Now, we are witnessing rapid development in banking industries to enable electronic payment through Internet as an example.

Although there have been significant advancement, the main problem remains the same, security and privacy. Banking industries play a significant role to ensure all financial transactions in digital form are adequately secured from any possible threats. However, there have been no single technical solutions available to handle all security issues in banking sector. It is even worsened if such security issues are regarded only from technical perspectives as confirmed in [1].

In the case of banking industries, better management of information security has been realized as an important factor to ensure safety of all financial transactions. Under IT management umbrella, we found several terms such as information technology governance, information security management, and information systems audit. In order to fulfill the requirements, banking industries follow several international standards to comply with, such as COBIT and ISO 27001.

The case study base on Indonesian banks which have implemented information security policy and audit systems based on COBIT or ISO 27001. COBIT or Control Objectives for Information and related Technology is a framework consists of a set of best practices for IT management with a subset of information security and assurance part [3]. Likewise, ISO 27001 is an international standard for information security management with best practice recommendations on information security management, as well as risks and controls within the context of an overall Information Security Management System (ISMS).

Deciding appropriate information security policy is not an easy task since there are many aspects should be considered appropriately. Therefore, there is a strong requirement to assist evaluation in this field.

We propose an evaluation method based on Analytic Hierarchy Process (AHP) which considering all relevant aspects of information security as a guidance framework. The following section describes the main concept of AHP. In section 3, we discuss two security cases of Indonesian banks. Then our analysis and discussion of the findings are provided in section 4. Finally, some concluding remarks are given at the end.

## II. ANALYTIC HIERARCHY PROCESS

Analytic Hierarchy Process (AHP) is originally introduced by Saaty in [4] as a excellent MCDM (multi criteria decision making) tool which was acknowledged by many researchers as can be seen in [8].

One of the main advantages of Saaty's AHP is it's simplicity compare toprevious decision support methods. It also enables qualitative and quantitative into the same decision making methodology by giving a basis for eliciting, discussing, recording, and evaluating the elements of a decision. It uses hierarchal way with goals, sub goals or factors and alternatives.

The structure will be then translated into a series of questions of the general form, '*How important is criterion A relative to criterion B*?'. The input to AHP models is the decision maker's answers to a series of questions is then termed pairwise comparisons. Questions of this type may be used to establish, within AHP, both weights for criteria and performance scores for options on the different criteria.

It is assumed that a set of criteria has already been established based on AHP model. For each pair of criteria, the decision-maker is then required to respond to a pairwise comparison question asking the relative importance of the two. Responses are gathered in verbal form and subsequently codified on a nine-point intensity scale [4][8] as follows:

TABLE I. AHP PAIRWISE COMPARISON VALUES

| How important is A relative to B? | Comparison Value |
|---|---|
| Equally important | 1 |
| weakly more important | 3 |
| strongly more important | 5 |
| very strongly more important | 7 |
| absolutely more important | 9 |

The value in between such as 2,4,6,8 are intermediate values that can be used to represent shades of judgement between those five basic assessments. If the judgment is that $B$ is more important than $A$, then the reciprocal of the relevant index value is assigned, for example if $B$ is considered to be strongly more important (5) than A as a criterion for the decision than $A$, then the value 1/5 (or 0.2) would be assigned to $A$ relative to $B$.

In some cases, judgments by the decision maker are assumed to be consistent in making decision about any one pair of criteria and since all criteria will always rank equally when compared to themselves, it is only ever necessary to make $1/2n(n-1)$ comparisons to establish the full set of pairwise judgments for $n$ criteria.

Then the results of all pairwise comparisons is stored in an input matrix $\mathbf{A} = [a_{ij}]$ that is an $n \times n$ matrix. The element $a_{ij}$ is the intensity of importance of criterion $n_i$ compared to criterion $n_j$. The following figure shows a typical matrix for establishing the relative importance of three criteria:

$$\begin{pmatrix} 1 & 3 & 5 \\ 1/3 & 1 & 7 \\ 1/5 & 1/7 & 1 \end{pmatrix}$$

Figure 1. AHP pairwise matrix.

In short, according to [8] one should follow four simple steps below in order to apply AHP method for guiding decision making process:
- Structure the problem into hierarchy.
- Comparing and obtaining the judgment matrix.
- Local weights and consistency of comparisons.
- Aggregation of weights across various levels to obtain the final weights of alternatives.

III. SECURITY ISSUES ON E-BANKING IN INDONESIA

The term electronic banking (or remote banking) is referred to the remotely conduct of traditional innovative banking activities with the use of electronic means [2].

In this section, two cases of internet banking security are discussed, BCA and Lippo Bank. The first case is BCA (Bank Central Asia) security incident in 2001.

The BCA case was basically known as "typo squatting" or URL hijacking. This type of attack relies on mistakes such as typographical errors made by Internet users when inputting a website address into a web browser. In this case, the attacker of BCA bought and managed several domain names (such as kilkbca.com kikbca.com, etc) slightly different to original one (klikbca.com). Then, all these fake websites were designed exactly the same with the original BCA website.

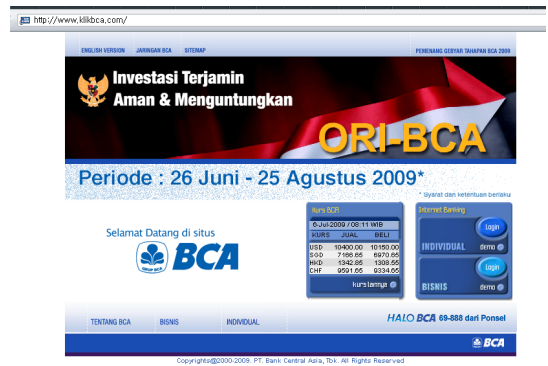

Figure 2. Original BCA website

This kind of attack exploit typographical errors made by BCA internet users. Then, for those who mistyped the BCA website, they were automatically directed to the fake website without realizing it since they saw exactly similar web presentation as the original BCA website.

This case obviously shows that internet banking still leave security holes that should not be underestimated by decision makers in banking industries.

The second example was LIPPO Bank case. It was in 2006 when several security professionals in Indonesia found and then reported security hole in LIPPO internet banking systems. The problem was coming from weakness on PIN distribution mechanism (see figure 3).

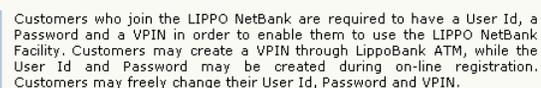

Figure 3. Security hole on LIPPO internet bank

Since customers may create their PIN (in this case it is called VPIN) through ATM machine, illegal persons may access internet access and change the number as reported in several stolen money through internet.

Based on both cases, we might see how information security policy plays a significant role to design proper internet banking service by considering all aspects.

## IV. ANALYSIS

This part describes the construction of AHP model, analysis of the model and result and discussion.

### A. Information Security Policy Model

In order to develop the model, first we classify information security related literatures into two main groups. First category is called information security aspect and the second one is information security elements.

Further, information security aspect can be classified into four main aspects, namely management, technology, economic and cultural aspects of information security as can be seen in the following table.

TABLE II. INFORMATION SECURITY ASPECTS

| Aspects | |
|---|---|
| Management | - *IT Governance*<br>- *Audit Information Systems*<br>- *Data classification*<br>- *Access control* |
| Technology | - *Software Security*<br>- *Network Security*<br>- *Internet Security* |
| Economy | - *Return of Security Investment*<br>- *Economic impact of security breaches* |
| Culture | - *Security awareness*<br>- *Security Education*<br>- *Organizational behavior* |

Managerial aspect of information security is one of several critical success factors of business organisation [9]. It covers strategic IT governance with emphasis in security and privacy and also evaluation in the form of IT auditing. As a result of its vital function, it is too risky to run a business without appropriate assurance for the security of its information systems operations [10].

Similarly, technological aspect of information security such as computer security [11], wired and wireless network security [12][13], and internet security [14], is a first consideration to develop secure information systems. This also can be seen from tremendous efforts to improve security quality by applying intrusion detection systems [15][16] and cryptography [17]. In short, technology is the critical point with respect to information security.

In terms of economy of information security, it is affirmed that economic considerations are important factor in recent information age [1][7] which should be included to as additional view point to strengthening information security.

Security investment is discussed in [18] to determine the optimal impact of such investment and its extensional effect [19] on information security.

Cultural aspect of information security represents the role changing culture in digital era and its relationship with security awareness through education [20]. It should become an embedded culture by individual within the organization [21].

Then, in terms of the second category which is information security elements, we suggest CIA which stands confidentiality, integrity and availability. It is applicable for our model since these triangle elements should become fundamental concern in all aspects mentioned before [23].

This is also due to widely recognition of CIA by security practitioners as three basic elements of information security that should be fulfilled in balance to guarantee appropriate security and privacy controls within an organization [22].

TABLE III. INFORMATION SECURITY ELEMENTS

| Elements | |
|---|---|
| Confidentiality | - *control disclosure of information*<br>- *authorize person or systems* |
| Integrity | - *data intact (no alteration)*<br>- *authorize person or systems* |
| Availability | - *data available and protected*<br>- *authorize person or systems* |

Table 3 represents the three security elements of CIA with specific attributes of each. Then, based on table 2 and 3, we develop information security policy evalustion model by following AHP structure as can be seen in figure 4.

### B. AHP Analysis

In this study, we use Web-HIPRE, a free applet based software to generate and analyze the AHP model [6].

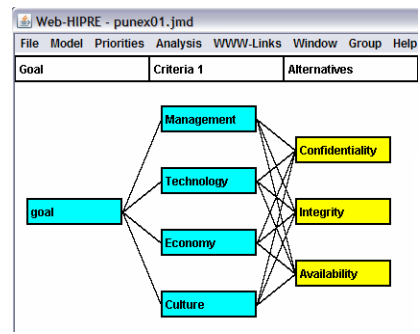

Figure 4. Structuring AHP model in Web-HIPRE

Fig. 4 shows the first AHP step to generate the information security policy model. Subsequently, all responses from respondents (CIO representatives of each banks) are put into the comparison window on each factor. The following figure shows qualitative result of composite overall priorities in Web-HIPRE.

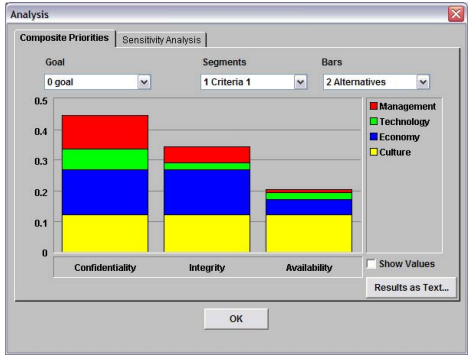

Figure 5. Overall AHP priorities

As can be seen, confidentiality is the highest priority among other two elements. Additionally, culture and economy are two important aspects of information security. The following table confirms the final result (quantitatively) of Fig 5 in more details.

TABLE IV. FINAL RESULT

|  | CONFIDENTIALITY | INTEGRITY | AVAILABILITY | TOTAL |
|---|---|---|---|---|
| MANAGEMENT | 0.112 | 0.054 | 0.011 | 0.177 |
| TECHNOLOGY | 0.068 | 0.023 | 0.023 | 0.114 |
| ECONOMY | 0.146 | 0.146 | 0.049 | 0.341 |
| CULTURE | 0.123 | 0.123 | 0.123 | 0.369 |
| TOTAL | 0.449 | 0.346 | 0.206 |  |

As mentioned earlier, confidentiality is the top priority consideration by decision makers in banking industries, which accounted for 0.449. It is followed by integrity and availability which both represent 0.346 and 0.206 respectively.

In addition, with respect to information security aspects, we found that decision makers in banking industries emphasize the importance of cultural and economy aspect with the value of 0.369 and 0.341. These are far higher that the last two aspects of management and technology which only accounted for 0.177 and 0.114 respectively.

C. Discussion

Information security with four main aspects has been accommodated properly through this model. As a significant industry in the country, banks are among the first mover institution to apply information technology in delivering the services.

Therefore, it is not surprisingly when we found that among other security elements, availability has least proportion rather confidentiality and integrity. It is found that decision makers in banks put more concern on confidentiality which accounted for 0.449 as the top priority. Integrity is the middle priority in banking industries which accounted for approximately 0.346 and availability is the last one with 0.206.

Confidentiality of financial data in banking industries has become a crucial point in order to prevent disclosure of information to unauthorized individuals or systems. Attacks in this area have been found in several reports which caused huge financial lost [24]. Similarly, banking industries also emphasize the important of data integrity by applying appropriate mechanism to guarantee that data cannot be modified without authorization only by authentic persons. Therefore, it is reasonable why these two elements of information security are highly appreciated in this sector.

Furthermore, in terms of four information security aspects, we found cultural aspect is the most important criteria among others which accounted for 0.369. The second priority is economy of 0.341, followed by management (0.341) and technological aspects which both accounted for 0.177 and 0.114.

Decision makers found that it is the time when culture in terms of behavior and education play more significant role in banking sectors. Previous cases also reflect this aspect on how important is security culture in cyber era. Costumers should be well informed on how to perform safe financial transaction on internet banking. In short it is reasonable for putting cultural aspect on the top priority among others.

As a core financial institution in the nation, banks put serious concern on economical aspect of information security threats. Security problems will potentially damage reputation of any banking industries. Lack of trust on banking systems will bring negative impact to economy. For that reason, economy is considered as the second priority.

In terms of managerial perspective, banking industries have been widely known with better management compare to other institutions. This is similar technology aspect which is the least portion found in this study. Banks have been recognized with more sophisticated information and computing technology since the beginning era of its development. Technological advancement was the focus of decision makers in the past. At the moment, it already has operational standard on how to operate and guarantee secure financial transaction from technical point of view.

V. CONCLUSION

Analytic Hierarchy Process can be used to help decision makers in banking sector analyzing information security policy from macro level perspective. This study justifies that the application of AHP method in information security is reasonable and it provides a robust and encompassing treatment for decision makers in both qualitative and quantitative ways.

From information security aspect perspective, the top priority is cultural aspect then followed by economy, management and technology respectively. Then, in terms of information security element, decision makers in baking industries emphasize the importance of confidentiality as the top consideration, followed by integrity as the middle priority and lastly availability.

ACKNOWLEDGMENT

The authors would like to thank ITPP Seoul National University for generous supports. Also, for anonymous reviewers for their valuable comments and suggestions, which are very helpful in improving the paper.

REFERENCES

[1] R. Anderson, "Why Information Security is Hard : An Economic Perspective", Proceedings of 17th Annual Computer Security Applications Conference 2001, pp. 10-14.
[2] Basle Committee, "Risk Management Principles for Electronic Banking" Basel Committee Publications, No. 98, July 2003, Bank for International Settlements.
[3] B. von Solms, "Information Security governance: COBIT or ISO 17799 or both?", Computers & Security vol. 24, issue 2, 2005, pp. 99-104
[4] T.L. Saaty, "The Analytic Hierarchy Process", RWS Publications, Pittsburgh, PA..1990
[5] J. Leiwo, C. Gamage, and Y. Zheng, "Organizational modeling for efficient specification of information security requirements", Advances in Databases and Information Systems: 3rd East European Conference, ADBIS'99, Maribor, 1999, pp.247-60.
[6] J. Mustajoki, and R.P. Hämäläinen,, "Web-HIPRE: Global decision support by value tree and AHP analysis", INFOR, vol. 38, no. 3, 2000, pp. 208-220
[7] S.E. Schecter, and D.S. Michael, "How much security is enough to stop a thief ? The economics of outsider theft via computer systems networks", Proceedings of the Financial Cryptography Conference, Guadeloupe. 2003, pp. 122-137.
[8] F. Zahedi, "The analytic hierarchy process—a survey of the method and its applications", Interfaces; vol.16, no. 4, 1986, pp. 96–108.
[9] R. Filipek, "Information security becomes a business priority", Internal Auditor, vol. 64, no.1, 2007 pp.18-21..
[10] M. Zviran, and W. Haga, "Password security: an empirical study", Journal of Management Information Systems, vol. 15 no.4, 1999, pp.161-85.
[11] C.E. Landwehr, "Formal Models for Computer Security", ACM Computing Surveys, vol. 13, issue 3, 1981, pp. 247-278
[12] S.D. Chi, J.S. Park, K.C. Jung, and J.S. Lee, "Network Security Modeling and Cyber Attack Simulation Methodology, in Information Security and Privacy", Lecture Notes in Computer Science, Springer Berlin / Heidelberg, 2001, pp. 320-333
[13] W.A. Arbaugh, N. Shankar, Y.C.J. Wan, and K. Zhang, "Your 80211 wireless network has no clothes", IEEE Wireless Communications, vol. 9, issue 6, 2002. pp. 44-51
[14] A. Householder, K. Houle, and C, Dougherty, "Computer attack trends challenge Internet security", Computer IEEE, vol. 35, issue 4, 2002, pp. 5-7.
[15] T. Bauss, "Intrusion detection systems and multisensor data fusion", Communications of the ACM, vol. 43, issue 4, 2000, pp. 99 - 105
[16] A. Fuchsberger, "Intrusion Detection Systems and Intrusion Prevention Systems", Information Security Technical Report, vol. 10, issue 3, 2005, pp. 134-139
[17] K.G. Paterson, "Cryptography from Pairings: A Snapshot of Current Research", Information Security Technical Report, vol. 7, issue 3, 2002, pp. 41-54
[18] L.A. Gordon, and M.P. Loeb, "The Economics of Investment in Information Security", ACM Transactions on Information and System Security, vol. 5, no. 4, 2002, pp. 438-457.
[19] L.A. Gordon,, M.P. Loeb, and W.Lucyshyn, "Sharing Information on Computer Systems Security: An Economic Analysis", Journal of Accounting and Public Policy, vol 22, no. 6. 2003, pp. 461-485
[20] M.E. Thomson, and R. von Solms, "Information security awareness: educating your users effectively", Information Management and Computer Security, vol. 6, no. 4, 1998, pp. 167–173.
[21] T. Schlienger, and S. Teufel, "Information Security Culture: The Socio-Cultural Dimension in Information Security Management", Proceedings of the IFIP TC11 17th International Conference on Information Security, 2002, pp. 191 - 202
[22] G. Dhillon, and J. Blackhouse, "Current directions in IS security research: towards socio-organizational perspectives", Information Systems Journal, vol. 11, no.2, 2001, pp.127-53.
[23] T. Peltier, Information Security Risk Analysis, Auerbach Publications, 2001 CRC Press, USA.
[24] CSI, CSI 2008 Survey, [Online document],[cited 2008 December 27] Available HTTP http://www.gocsi.com